\documentclass[a4paper]{article}

\usepackage[T1]{fontenc}
\usepackage[utf8]{inputenc}
\usepackage{amsmath}
\usepackage{cases}
\usepackage{amsthm}
\usepackage{amsfonts}
\usepackage{mathtools}
\usepackage{tikz}
\usepackage{pgf}
\usepackage{graphicx}
\usepackage{framed}
\usepackage{multicol}
\usepackage{todonotes}
\usepackage{float}
\usepackage{wrapfig}
\usepackage{pdflscape}
\usepackage{rotating}

\usepackage{hyperref}
\usetikzlibrary{arrows,automata}
\usetikzlibrary{positioning,calc,decorations.pathmorphing,decorations.pathreplacing,patterns}

\newcommand{\citep}[1]{%
\cite{#1}}
\usepackage{todonotes}

\newcommand{\closeopenc}[0]{\end{framed}}

\newcommand{\openopenc}[0]{\begin{framed}
\vspace{-.5cm}
\paragraph{In reconciliation services}
}

\newcommand{\closeopenr}[0]{\end{framed}}

\newcommand{\openopenr}[0]{\begin{framed}
\vspace{-.5cm}
\paragraph{In OpenRefine}
}

\usepackage{textcomp}

\title{A survey of OpenRefine reconciliation services}
\author{Antonin Delpeuch \\
  Department of Computer Science \\
  University of Oxford, UK \\
\small{\href{https://orcid.org/0000-0002-8612-8827}{\includegraphics[scale=0.55]{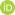} \texttt{0000-0002-8612-8827}}}}
\date{\today}

\begin{document}

\maketitle

\begin{abstract}
  We review the services implementing the OpenRefine reconciliation API,
  comparing their design to the state of the art in record linkage.
  Due to the design of the API, the matching scores returned by the services
  are of little help to guide matching decisions.
  This suggests possible improvements to the specifications of the API,
  which could improve user workflows by giving more control over the scoring
  mechanism to the client.
\end{abstract}

\section*{Introduction}

Integrating data from sources which do not share common unique
identifiers often requires matching (or \emph{reconciling},
\emph{merging}) records which refer to the same entities.
This problem has been extensively studied and many heuristics have been
proposed to tackle it~\citep{christen2012data}.

The OpenRefine reconciliation API\footnote{\url{https://github.com/OpenRefine/OpenRefine/wiki/Reconciliation-Service-API}} is a web protocol designed for this task,
which was initially implemented by the Freebase knowledge base. While most
software packages for record linkage assume that the entire data is available
locally, and can be indexed and queried at will, this API proposes a workflow
for the case where one of the data souces to be matched is held in an online
database. By implementing such an interface, the database lets users match their
own datasets (which are typically smaller in size) to the identifiers
it holds.

As entity matching often relies on names, the reconciliation API is
essentially a search API tailored to the reconciliation problem. A
typical query to a reconciliation interface consists of the name of
the entity to search for, an entity type to restrict the search to a
certain category of entities and a couple of other attributes to
refine the search by field values. The service responds by returning
matching candidates with their identifiers.

The canonical client for this API is
OpenRefine\footnote{\url{http://openrefine.org/}}~\citep{openrefine}, an
Extract-Transform-Load tool which can be used to transform raw tabular
data into linked data. The tool proposes a semi-automatic approach to
reconciliation, making it possible for the user to review the quality
of the reconciliation candidates returned by the service.  To that
end, the reconcilation API lets services expose auto-complete
endpoints and HTML previews for the entities they store, easing
integration in the user interface of the client.

In this survey, we review the current ecosystem of reconciliation
services.  We analyze how they use the various features of the
reconciliation API, review their underlying implementation when
available, and propose possible changes to the protocol, making it
more useful to data providers and consumers.

\subsection*{Acknowledgements}

We thank the OpenCorporates team and Vladimir Alexiev for their
feedback on this project.
This work was supported by
\href{https://opencorporates.com/}{OpenCorporates} as part of the
``TheyBuyForYou'' project on EU procurement data.  This project has
received funding from the European Commission's Horizon 2020 research
and innovation programme (grant agreement n\textdegree 780247).

\section{Overview of reconciliation}

We first explain what reconciliation means and how
the OpenRefine reconciliation API can be used for this process.

\subsection{Goals and scope}

Reconciliation consists in establishing a mapping between two
sets of entities:
\begin{itemize}
\item the entries of a dataset $U$ provided by the user.  Such a
  dataset would typically contain a few hundreds or thousands of
  entries. For instance, the dataset could list procurement contracts
  between some administration and its suppliers, or a list of
  endangered plants in a national park, or an inventory of coins found
  on an archeological site.
\item the records of an authoritative online database $D$. This
  database is typically larger and considered more reliable than the
  user dataset and generally contains unique identifiers for its
  entities. For instance,
  \href{https://opencorporates.com/}{OpenCorporates} lists about 165
  million companies harvested from company registers, with their
  official identifiers, the \href{http://www.ipni.org/}{International
    Plant Names Index} stores canonical plant names and the associated
  scholarly information, and \href{http://nomisma.org/}{Nomisma}
  curates linked open data around numismatics.
\end{itemize}

The goal of reconciliation is to guess a partial function $f : U \to
D$, mapping each entry from the user dataset to the record in $D$ that
represents the same entity, if any. This matching process is therefore
bipartite: it is assumed that both $U$ and $D$ are free from
duplicates and records coming from the same database are not compared
to each other. The function $f$ is partial: it is possible that a user
record does not correspond to any reference entry in $D$. Each user
record corresponds to at most one entity in $D$.

The two databases $U$ and $D$ generally contain different fields. For
reconciliation to be possible, we assume that some of them are shared
by both databases. For instance, it is often the case that both
databases contain names for the entities they refer to. The mapping
$f$ is then constructed by comparing the values of these common
attributes.  These values generally differ slightly in both databases,
and are ambiguous or incomplete, which is why heuristics have to be
used to construct $f$.

The motivations for reconciliation vary. One can reconcile to make the
user dataset more canonical, by normalizing the references to the
entities to match the authoritative data. One can also enrich the user
data with additional identifiers and other attributes retrieved from
the target database. Finally, it is also possible to use
reconciliation as part of the curation process of the authoritative
database, for instance to push data from $U$ into $D$.

\subsection{The OpenRefine reconciliation API}

In this section we give an overview of the specifications of the reconciliation API.
We are not aware of any formal specification for it, but an informal description
can be found at \url{https://github.com/OpenRefine/OpenRefine/wiki/Reconciliation-Service-API}.
We start by formalizing the data model projected by the API on the reconciliable data source.

\paragraph{Data model}
The data source $D$ is assumed to have a set of entities $E$, a set of
possible entity types $T$ and a set of possible properties $P$. Types
provide a way to categorize entities, and properties are predicates
that can be applied to entities. We will also denote by $\Sigma^*$
the set of character strings.

Each entity $e \in E$ has a set of types $\text{types}(e) \subseteq
T$, possibly empty. For each property $p \in P$, there is a set of
values $\text{eval}(e, p) \subseteq V = \Sigma^* \cup E$, possibly
empty. A value $v \in V$ can either be a character string $s$ or
another entity $e' \in E$.

For instance, if the data source is structured as a relational
database, then types could be tables, entities could be table rows and
properties could be table columns. Each entity would have exactly one
type (the table the row belongs to) and $\text{eval}(e, p)$ would be
empty (if the row and the column do not belong to the same table) or
a singleton (the value in the intersection of the row and column).

If the data source is structured as a graph database, then entities
could be nodes, types could be some sort of category system on nodes, and
properties could be graph edges.

Each entity, type and property is designated by an identifier
$\text{id}(x) \in \Sigma^*$. It also has a human-readable name
$\text{name}(x) \in \Sigma^*$.

\paragraph{Reconciliation queries}

A reconciliation query $q$ is given by:
\begin{itemize}
\item a query string $s \in \Sigma^*$, representing the name of the entity queried for;
\item an optional set of types $t \subseteq T$ to restrict the search to;
\item a set $m \subseteq P \times V$ of property values, possibly empty, giving additional
  information about the entity queried for.
\end{itemize}

The main task of a reconciliation service is to process reconciliation queries,
sent over HTTP, and return a set of matching candidates for each query.

A reconciliation candidate $c$ is given by:
\begin{itemize}
\item an entity $e \in E$, serialized with its id, name and types;
\item a matching score $x \in \mathbb{R}$. The definition of this score is left to the service,
  but it is expected that the higher it is, the better the candidate matches the query.
\end{itemize}

If we denote by $Q$ the set of reconcilation queries and $C$ the set
of reconciliation candidates, the task of a reconciliation service is
therefore to compute a function $Q \rightarrow \mathcal{P}(C)$.

\paragraph{Suggest, preview and extend services}
In addition to the main querying method, the reconciliation API also
specifies ways for the service to expose suggest (or auto-complete)
services, HTML previews of entities and bulk data retrieval, which
ease the integration of the service in OpenRefine's user interface.

\section{Analysis of existing reconciliation services}

In this section, we survey the reconciliation services currently accessible
online, and outline the main technical choices behind them. We compare
them to the general techniques found in the literature on record linkage.
Although there are many different approaches to record linkage,
most of them broadly follow a common architecture \cite{elmagarmid2007duplicate,christen2012data}.
Figure~\ref{fig:services-table} provides a summary of the characteristics
of the reconciliation services studied.

\paragraph{Candidate retrieval}
First, potential matches from the target database $D$ are selected. It
is generally assumed that the user data contains a name for the
entity. In the case of companies, this name would ideally be the
official company name, but could also be an acronym, a trademark or
any other name under which the company is or was known
informally. This name is generally the primary discriminative
information at this stage, if not the only one. A short list of
candidates with similar names are retrieved from the target database,
generally using a search index. The approaches for this step are
reviewed in Section~\ref{sec:candidate-retrieval}.

\paragraph{Field scoring}
Second, each field supplied by the user is compared against the
corresponding values in the target database. The degree of similarity
of each value pairs is usually represented by a boolean or numeric
score. The nature of the scoring method depends largely on the type of
information stored in each field. In addition, some entities from the
database can be assumed to be more prevalent or popular than others,
so a popularity score can be computed independently of the data
supplied in the query. In Section~\ref{sec:field-scoring}, we survey
field and popularity scoring techniques and their use in
reconciliation services.

\paragraph{Matching decision}
Third, the field scores are used to determine which of the candidates
(if any) will be retained as the matching entity. This critical step
is often broken down into two tasks: first, aggregate all the
field-level scores into one global matching score for each candidate
in the short list. The crucial decision at this stage is to balance
the influence of each field on the final matching score. In
Section~\ref{sec:global-scoring-methods} we give an overview of the
wide range of approaches that have been proposed to determine these
weights. Once this global score is defined, the final decision whether
to match a candidate to the user data is generally based on a
threshold for the global score, determined by the risk associated with
false positives and false negatives.

\subsection{Candidate retrieval} \label{sec:candidate-retrieval}

The choice of restricting the matching heuristics to a short list of
candidates is a simplification to reduce the computational cost of
reconciliation. Instead of comparing the user record to each database
entry, these comparisons are restricted to the most relevant entries,
fetched by a coarse-grained but computationally efficient filtering
method.

The usual way to perform this filtering is by maintaining an index on
one or more fields of the database. This indexing is often called
\emph{blocking} \citep{christen2012survey,christen2012data}: the records are
partitioned into blocks (or buckets) depending on their values. Given
a query, we compute the corresponding block value (or multiple blocks)
and only retrieve candidates from these blocks. For instance, indexing
names with a phonetic transcription such as Soundex will map the names
``Will'' and ``Wil'' to the same code \texttt{W400}.

Common solutions involve building an inverted index which can be used
to retrieve all candidates containing words in the
query~\cite{2013elasticsearch}. Various techniques have been developed
to make this retrieval more error-tolerant: for instance, indexing
based on q-grams (sequences of $q$ characters) instead of words makes
it possible to retrieve misspelt records
\citep{baxter2003comparison,christen2012data}.

\openopenc

Existing services overwhelmingly rely on traditional search engines
for candidate retrieval, and a Lucene-based index is the most common
choice (both ElasticSearch and Solr rely on Lucene). This holds both
for services hosted by the original data provider, which can query
their own search engine directly and for services implemented by
third-parties on top of the generic API exposed by the data provider.

Many of the advanced indexing techniques
and linguistic preprocessing mentioned in the literature are available
in Lucene.  In this context, improving this candidate retrieval step
consists in tuning the configuration of the indices to the type of
data they are used for. It is an area worth investing effort as any
improvement benefits not only the reconciliation service
but also all the other services relying on the search engine (seach as
any search UI proposed to end users).

The only exceptions to this are the GODOT reconciliation service,
where no candidate retrieval phase is done (all records are compared to the query),
and OpenCorporates where some queries rely on SQL search.
\closeopenc

\subsection{Individual field scoring} \label{sec:field-scoring}

In this section, we review various scoring methods for individual
fields.
 
In the absence of unique identifiers, the name of an entity is the
primary discriminative clue to identifiy it. Therefore, scoring methods
for entity names have attracted a lot of attention.  They fall into
three families depending on which basic comparison unit they use:
characters, q-grams or words.

Character-based metrics quantify the minimal number of operations on
individual characters to transform a string into the string it is
compared to~\cite{christen2012data}. The nature and cost of the
operations involved depends on the algorithm. The simplest version is
called the edit distance: the allowed operations consist in deleting,
inserting or replacing characters. Although the search space of
editing operations is large, it is possible to compute this distance
quickly with the Levenshtein algorithm: the number of operations is
proportional to the product of the lengths of the strings
compared. Many variants of this metric have been introduced: adding
operations to modify larger groups of consecutive characters (for
instance to soften the effect of a missing word or shortened word on
the score), giving different weights depending on the characters
substituted (to account for replacements of similar characters such as
O and 0)~\cite{needleman1970general} or speeding up the comparison by
restricting the number of changes~\cite{landau1989fast}.

Q-gram based metrics compute all the sequences of $q$ consecutive
characters in each string, and compare them. For instance, the word
``Oracle'' contains the 3-grams ``Ora'', ``rac'', ``acl'' and
``cle''. Although not as precise as an edit distance, comparing the
sets of q-grams contained in strings is a simple and inexpensive way
to assess to which extent they differ. It accounts for word
reorderings and can also be used for indexing. Like character-based
metrics, q-grams are mostly useful when the strings differ by spelling
mistakes or encoding errors.

Word-based (or token-based) distances first separate the input into
words and use these as basic comparison units. Working at word level
gives a more semantic notion of similarity, without conflating words
with similar spellings but unrelated meanings. It is still possible to
add stemming and other normalization procedures to each word to account
for some controlled variation in word spellings.

For both Q-gram and word-based approaches, there are various methods
to turn a set of common units into a score. The simplest way is to
count the number of matching units and divide it by the total number
of units in both strings, for instance. However, not all tokens in a
name are equally informative. For instance, the similarity between
``Greentech Distribution'' and ``Greentech Services'' should be
higher than that of ``Greentech Distribution'' and ``Globafrik
  Distribution'', simply because having ``Greentech'' in common is
more informative than ``Distribution''.

The standard solution to this problem is called TF-IDF (Term Frequency
- Inverse Document Frequency). Informally, this is a method to measure
the significance of a word occurence in some text. The significance is
proportional to two factors: how often the term appears in the given
document and how rarely it appears in general in other documents. In
the context of name matching, the documents are very short as they are
the names themselves, so term frequency does not play an important
role.  However, inverse document frequency is a decisive factor which
will give more significance to ``Greentech'' than to ``Distribution''.
SoftTFIDF~\citep{cohen2003comparison} is a method to use TF-IDF as a
string similarity measure. In its simplest version, it is simply
defined as
$$\text{SoftTFIDF}(A,B) = \sum_{w \in A \cap B} t(w,A)t(w,B)$$ where
$A, B$ are the strings to compare, $w$ ranges over the common words in
$A$ and $B$, and $t(w,A)$ is the TF-IDF weight of $w$ in $A$.  In its
full version, SoftTFIDF also allows for some dissimilarity between
words by incorporating a word similarity metric in each summand.

Popularity scores can help introduce a bias towards more sailent
entities in the database. Their nature largely relies on the type of entities
at hand. They can be based on particular data fields of the entities
(such as company revenue or number of employees), or on the interlinking
structure between them (PageRank).

\openopenc
Many reconciliation services delegate field and popularity scoring
to their underlying search engine. Their role therefore amounts to translating
reconciliation queries to corresponding search queries, crafted to
obtain the desired scoring. Text fields are therefore generally scored using
variants of TF-IDF. Few reconciliation services
include popularity metrics.

The reconciliation services that do compute a
matching score for the name or other textual fields generally use
Levenshtein-based metrics or more conservative exact matching.
\closeopenc

\subsection{Global scoring methods} \label{sec:global-scoring-methods}

Once fields from the reference database have been compared with user
data, we need to draw on these comparisons to decide whether to match
the user record to a reconciliation candidate. Users have various
expectations about this step and it is crucial to accomodate
them.

First, reconciliation is an inherently approximate process and the
accuracy to aim for depends on the application: the cost of false
positives (erroneously matching a user record to a reference
identifier) and false negatives (erroneously declaring that the user
record does not correspond to any reference identifier) varies.  Many
record linkage methods let the user influence these error rates by
computing a global matching score. The user can then set their own
threshold on the matching score to get the desired trade-off between
false positives and false negatives. However, in the absence of
reference data to evaluate these error rates, the impact of the
threshold on errors is often unknown.

Second, the notion of identity between the user data and the reference
database is not always the same. For instance, when reconciling
companies from a list of bids for a market, users might want to match
each company to the exact legal entity who submitted the bid, or to a
better known larger entity controlling the bidder. This means that the
relative influence of fields such as the headquarters' location might
also need to vary. Giving the user some control over the global
scoring method is also important to let them factor in the reliability
of their data in each field.

Given a collection of features comparing a user record to a reference
record, there are various ways to build a decision function which
predicts whether the records refer to the same entity.

\paragraph{Linear models}
Features are often boolean or numeric, and the simplest way to combine
them into one score is to use a linear combination of the
features. The higher this weighted sum gets, the more confident the
system is that the two records represent the same entity. Many
probabilistic approaches to record linkage such as that of Fellegi and
Sunter~\cite{fellegi1969theory} also fall into this category: the
score corresponds to a probability of match and it is log-linear in
the feature values. The decision whether two records are considered as
matches or mismatches is then taken by comparing the confidence score
to thresholds. In probabilistic models these thresholds can be
determined by desired false-positive and false-negative
rates~\citep{elmagarmid2007duplicate}.

\paragraph{Decision trees}
Decision trees define simple decision procedures to decide whether two
records match, without implicitly defining a global matching score~\citep{cochinwala2001efficient}.
Starting from the root of the tree, the decision process visits nodes.
Each internal node is associated with one feature and a threshold to
compare it against. The comparison determines which node to visit
next, and the process terminates when a leaf is visited, which
contains a matching decision. One important aspect of decision trees is
that they are easy to define and interpret for users.

\paragraph{Other classifiers}
Deciding if two records refer to the same entity is a binary
classification problem so many other classes of decision functions can
be used to tackle this problem. Generic machine learning tools such as
Support Vector Machines or K Nearest Neighbours have been used in this
context~\citep{christen2008automatic,christen2012data}.

\openopenc
Again, many reconciliation services avoid developing their own scoring
mechanisms by simply exposing the score exposed by the underlying search
engine. When the service is run by the data provider itself, the configuration
of the search index can be adjusted to make this score more useful.

When scoring is done explicitly in the service, linear models are the
most widespread choices in reconciliation services, due to their
simplicity and their ability to aggregate evidence from various
features. However, given the partial view reconciliation services have
on the user data, a probabilistic approach is difficult, making it
hard to set weights and thresholds in a principled way.

When matching or unmatching sets of rows selected by
facets, OpenRefine users are effectively building an implicit decision
tree in their operations history. However, given that the field
matching scores are not exposed to the user, this work often involves
re-computing locally similar features (such as edit distances between
labels). See Section~\ref{sec:exposing-field-scores} for proposals to
improve this.

We are not aware of any use of advanced machine learning
techniques in combination with OpenRefine or its reconciliation
API. The limiting factor for this is again the unavailability of field
matching scores, which we also propose to solve in
Section~\ref{sec:exposing-field-scores}.  \closeopenc

\section{Improving reconciliation workflows} \label{sec:api-workflows}

OpenRefine reconciliation is designed to solve a particular form of
record linkage problem. It was originally designed to work with
Freebase, a collaborative knowledge graph. In this context, users
would align datasets that they want to upload to Freebase by matching
the entities in their table to existing Freebase topics, so that the
information that they upload builds up on previous contributions by
improving existing topics and creating new ones. The reconciliation
process has then been generalized to work with arbitrary target
databases, by specifying a dedicated web API that the database must
expose \citep{2018reconciliation}.

In this section, we describe what the current
reconciliation workflow looks like, what its limitations are, and how
it could evolve to accomodate better for users' needs.

\subsection{Current reconciliation workflow} \label{sec:current-workflow}

\begin{figure}
  \centering
  \includegraphics[scale=0.4]{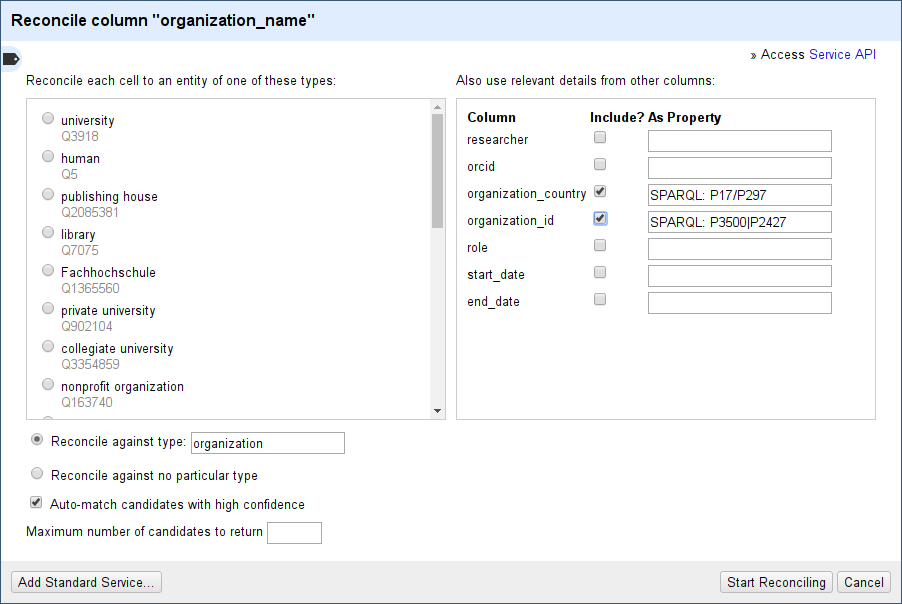}
  \label{fig:openrefine-ui}
  \caption{OpenRefine's user interface for reconciliation configuration}
\end{figure}

OpenRefine lets users link their tables to target databases such as
OpenCorporates.  This works by selecting a column, containing names of
the entities to match, and configuring the reconciliation process as
shown in Figure~\ref{fig:openrefine-ui}.

Users can choose to restrict the reconciliation candidates to records
of a particular type. This notion of type is defined by the
target database, each record they hold can have multiple types, each
of which is defined by an identifier and accompanied by a
human-readable name.  Beyond the column containing names, it is
possible to use other columns of the table by matching it to fields of
the target database. To this end, the target database must expose a
vocabulary which lists the fields that user data can be matched against.

Once reconciliation has been configured, OpenRefine will make a series
of API calls to the reconciliation service, each call containing a
small batch of reconciliation queries. A reconciliation query consists
of the values in the columns used for reconciliation (main column for
the name and auxiliary columns for other fields) as well as the chosen
type to restrict reconciliation candidates to (if any).  For each
reconciliation query, the reconciliation service returns a list of
candidates. Each candidate is supplied with a unique identifier from
the database, a human-readable name, a list of types and a matching
score.

This matching score is produced by the reconciliation service on the
basis of the information supplied in the query and is typically opaque
- users do not necessarily know how scoring works. In particular,
users do not have any easy way to influence the importance of a given
field, or to inspect the matching scores of individual fields.

By using facets, it is then possible to take matching decisions for
rows matching certain criteria. These criteria can depend on matching
scores, types of the candidates, or any other value in the table.

\subsection{Exposing field-level scores in the reconciliation API} \label{sec:exposing-field-scores}

The main limitation of this workflow is the lack of control on the
scoring mechanism. As a user, it is hard to rely on an opaque score to
create a reliable reconciliation workflow. Even if the scoring
function is publicly documented, it might not be suitable for all
datasets. As a reconciliation service provider, coming up with a
scoring mechanism that works for everyone is impossible, especially
because the final matching decisions made by users are not
communicated to the reconciliation services: it is impossible to learn
the scoring function from data unless a dedicated dataset is annotated
separately. Such a dataset is hard to compile given the wide variety
of use cases and user data sources that reconciliation endpoints are
typically exposed to.

\begin{figure}
  \hspace{-2cm}
  \begin{tikzpicture}[xscale=3]
    \foreach \x/\txt in {1/Candidate retrieval,2/Field\\ scoring,3/Global\\ scoring,4/Matching decision} {
      \node[text width=2cm,align=center,rectangle,draw=gray] at (\x,1) (b\x) {\txt};
    }
    \draw[-latex] (b1) -- (b2);
    \draw[-latex] (b2) -- (b3);
    \draw[-latex] (b3) -- (b4);
    \draw[-latex] ($(b1)+(-.7,0)$) -- (b1);
    \draw[-latex] (b4) -- ($(b4)+(.7,0)$);
    \foreach \x/\txt in {1/Offline matching,2/Manual matching\\ via search API,3/Proposed new\\ reconciliation API,4/Current OpenRefine\\ reconciliation API,5/Server-side dataset\\ matching} {
      \node[circle,draw] at ($(\x,0)-(.5,0)$) (c\x) {\x};
      \draw[dashed,gray] (c\x) -- +(0,2);
      \node[node distance=1.4cm,below of=c\x,text width=2.7cm,align=center] {\txt};
    }
    \node at (0.1,1.8) {Server side};
    \node at (4.9,1.8) {Client side};
  \end{tikzpicture}
  \label{fig:responsibilities}
  \caption{Five possible boundaries of responsibilities between server and client in a reconciliation process}
\end{figure}
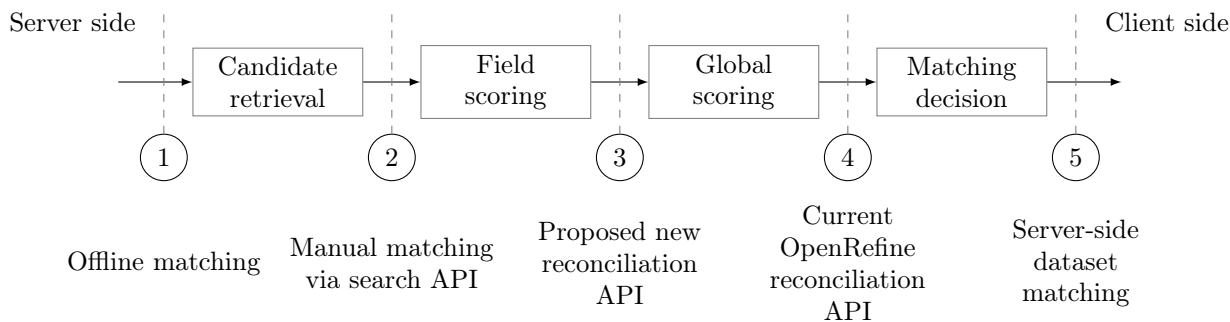

To solve this problem, we need to shift the boundary between the
responsibilities of the service provider and the user in the
reconciliation process. Figure~\ref{fig:responsibilities} shows a
diagrammatic representation of the reconciliation process, with
various options as to where the reconciliation API should sit. Dashed
lines represent the separation of responsibilities between client on the right (the
user who supplies the data to match) and the server on the left (the
reconciliation service which exposes the database to be matched
against) in various scenarios. Each of these scenarios has important
implications in terms of usability, performance and quality that we
analyze below.

\paragraph{1 Offline matching} This consists in downloading a copy
of the target database and performing the reconciliation process
locally. It is generally necessary to build indices on the database
first, transform it to a different format, and write some custom
matching heuristics.  Off-the-shelf record linkage tools such as
Duke\footnote{\url{https://github.com/larsga/Duke}}, the R package
RecordLinkage~\citep{sariyar2010recordlinkage} or
Serf~\citep{benjelloun2009swoosh} can also be used. This workflow can
be necessary when the user dataset to be matched is large, as it
minimizes data transfer between the user and the database. However, for
users with small datasets or simple matching needs this workflow is
completely impractical when the target database is large, as in the case of OpenCorporates.

\paragraph{2 Manual matching via search API} This workflow is fairly widespread,
as many online databases offer a web API that can be used to search
for records using various criteria. It is then up to the user to
decide how to compare the records returned to their own data. If the
API exposed by the service is flexible enough to retrieve
the appropriate candidates efficiently, this can be viable but a
custom reconciliation process must be implemented by the user, which
is a significant investment. The API often hides valuable statistics
from the search engine, such as those needed to compute TF-IDF scores.

\paragraph{3 Proposed new reconciliation API} We propose to improve the existing
reconciliation API used in OpenRefine to let data providers expose
matching scores for individual fields, instead of just one global
score. This would let clients use their own global scoring methods,
which would give the appropriate weight to each field. Handling field
scoring server-side would let the reconciliation services implement
metrics that are meaningful in their domain (such as the bespoke name
matching heuristics used in OpenCorporates for company names) without
having to implement this domain-specific expert knowledge in generic
client-side tools. In this configuration, field-level scores can also
depend on statistics maintained in the search engine of the database,
making it possible to use TF-IDF scores for instance. We explore the
implications of this proposal in
Section~\ref{sec:new-reconciliation-API}.

\paragraph{4 Current OpenRefine reconciliation API} As explained in
Section~\ref{sec:current-workflow}, global candidate scoring is
currently the responsibility of the reconciliation service, making it
impossible for users to influence how this score is computed. Another
way to solve this problem would be to let the user specify more
parameters in their reconciliation queries, such as providing a
numerical weight factor for each of the columns supplied. The main
downside with such an approach is that it would be hard for the user
to come up with these weights initially, and changing the weights a
posteriori would require running again the reconciliation process
(which costs time and resources). This would make it hard to integrate
the API with any machine learning approach.

\paragraph{5 Server-side dataset matching} In this scenario, the user would
upload their dataset to the reconciliation service, which would
perform the matching of all rows in one go and return the final
results. This would have the advantage of eliminating round-trips
between the client and server, but would make it hard for the user to
finely tune the reconciliation heuristics.  Providing reference
matching decisions to the reconciliation service would also be hard as
the reconciliation candidates would not be known in advance.

\subsection{Evolution of the reconciliation API} \label{sec:new-reconciliation-API}

As motivated by our analysis of the various scenarios above, we
propose changes to the reconciliation API and evaluate the impact
on service providers, API clients and end users.

For a service provider, the proposed change would imply changing the
format of the responses returned to include the matching scores for
each field.  The specification of the API could potentially allow for
multiple scores per field, which would let services expose different
scoring heuristics. The different scores returned could also be
independent from the fields supplied: the API would require services
to return an arbitrary list of feature values. In order to be
compatible with existing clients, it might be useful to require the
services to still return a global score as well. This global score
would serve as a default and could be ignored by clients which can
rely on the individual features instead. These details and the
concrete format of these responses should be proposed for consultation
with the community to ensure that the initiative is followed by as
many stakeholders as possible.

For an API client, such as OpenRefine itself, this proposal would
imply some changes to at least store and expose the feature scores.
OpenRefine already has a dedicated field to store features associated
with a particular reconciliation candidate but these features are
computed locally and are therefore very generic and not easily
adaptable to particular domains. More importantly, clients need to
include tools to help users build appropriate decision functions for
their dataset. This could be achieved by integrating machine learning
packages developed in other tools, which would give the user a real
control over the error rates and abstract away the features. The
existing manual matching capabilities could be reused to provide
training data to these automated approaches.  Active learning has been
applied to record linkage to learn classifiers with small quantities
of training
data~\citep{tejada2002learning,sarawagi2002interactive}. Active
learning works by incrementally improving a classifier with new
training examples, selected from cases where the classifier has the least
confidence. This learning paradigm can be used with a wide range of
types of classifiers~\citep{arasu2010active} and could be an
interesting complement to the exploratory data analysis workflows
encouraged by OpenRefine's design.

For users, the reconciliation process must remain accessible and
simple. It should be possible to work with a stock global scoring
method whose performance should be comparable to the current scores.
Exploring the values of these features should be possible with facets,
and features should be documented so that users can understand the
meaning of reconciliation results. Finally, in a scenario where
machine learning is used, it should ideally be possible to expose the
learned decision function to the user, for instance as a decision tree
or by drawing the decision boundary on a scatterplot. It could also be
useful to let the user interact with this learned model and tune it
with their own knowledge of the data.

\section*{Conclusion}

We have surveyed the existing reconciliation services and compared
them to the state of the art in recorcd linkage.  From this review, we
suggest possible changes to the reconciliation API. We propose to make
it possible for reconciliation services to expose field matching
scores in addition to (or instead of) global matching scores for each
candidate. The initial motivation for this change is to make it
possible for users to balance the importance of each field, but the
implications are much broader as this change would make it possible to
reuse a wide range of advanced classifiers from the literature. With
the appropriate integration in OpenRefine, this would help users build
reliable matching heuristics, informed by their expert knowledge of
the data.  This change would also benefit any other API user who could
feed these features to the machine learning packages of their choice.

\bibliographystyle{plain} \bibliography{zotero}

\begin{thebibliography}{10}

\bibitem{2013elasticsearch}
Elasticsearch from the {{Bottom Up}}, {{Part}} 1.
\newblock https://www.elastic.co/blog/found-elasticsearch-from-the-bottom-up,
  September 2013.

\bibitem{2018reconciliation}
Reconciliation {{Service API}}.
\newblock https://github.com/OpenRefine/OpenRefine, November 2018.

\bibitem{arasu2010active}
Arvind Arasu, Michaela G{\"o}tz, and Raghav Kaushik.
\newblock On active learning of record matching packages.
\newblock In {\em Proceedings of the 2010 International Conference on
  {{Management}} of Data - {{SIGMOD}} '10}, page 783, {Indianapolis, Indiana,
  USA}, 2010. {ACM Press}.

\bibitem{baxter2003comparison}
Rohan Baxter, Peter Christen, and Tim Churches.
\newblock A {{Comparison}} of {{Fast Blocking Methods}} for {{Record Linkage}}.
\newblock page~6, 2003.

\bibitem{benjelloun2009swoosh}
Omar Benjelloun, Hector {Garcia-Molina}, David Menestrina, Qi~Su,
  Steven~Euijong Whang, and Jennifer Widom.
\newblock Swoosh: A generic approach to entity resolution.
\newblock {\em The VLDB Journal}, 18(1):255--276, January 2009.

\bibitem{christen2008automatic}
Peter Christen.
\newblock Automatic record linkage using seeded nearest neighbour and support
  vector machine classification.
\newblock In {\em Proceeding of the 14th {{ACM SIGKDD}} International
  Conference on {{Knowledge}} Discovery and Data Mining - {{KDD}} 08}, page
  151, {Las Vegas, Nevada, USA}, 2008. {ACM Press}.

\bibitem{christen2012data}
Peter Christen.
\newblock {\em Data Matching: Concepts and Techniques for Record Linkage,
  Entity Resolution, and Duplicate Detection}.
\newblock {Springer Science \& Business Media}, 2012.

\bibitem{christen2012survey}
Peter Christen.
\newblock A {{Survey}} of {{Indexing Techniques}} for {{Scalable Record
  Linkage}} and {{Deduplication}}.
\newblock {\em IEEE Transactions on Knowledge and Data Engineering},
  24(9):1537--1555, September 2012.

\bibitem{cochinwala2001efficient}
Munir Cochinwala, Verghese Kurien, Gail Lalk, and Dennis Shasha.
\newblock Efficient data reconciliation.
\newblock {\em Information Sciences}, 137(1):1--15, September 2001.

\bibitem{cohen2003comparison}
William~W Cohen, Pradeep Ravikumar, and Stephen~E Fienberg.
\newblock A {{Comparison}} of {{String Metrics}} for {{Matching Names}} and
  {{Records}}.
\newblock page~6, 2003.

\bibitem{elmagarmid2007duplicate}
Ahmed~K. Elmagarmid, Panagiotis~G. Ipeirotis, and Vassilios~S. Verykios.
\newblock Duplicate {{Record Detection}}: {{A Survey}}.
\newblock {\em IEEE Transactions on Knowledge and Data Engineering},
  19(1):1--16, January 2007.

\bibitem{fellegi1969theory}
Ivan~P Fellegi and Alan~B Sunter.
\newblock A theory for record linkage.
\newblock {\em Journal of the American Statistical Association},
  64(328):1183--1210, 1969.

\bibitem{openrefine}
David Huynh, Tom Morris, Stefano Mazzocchi, Iain Sproat, Martin Magdinier, Thad
  Guidry, Jesus~M. Castagnetto, James Home, Cora {Johnson-Roberson}, Will
  Moffat, Pablo Moyano, David Leoni, {Peilonghui}, Rudy Alvarez, {Vishal
  Talwar}, Scott Wiedemann, Mateja Verlic, Antonin Delpeuch, {Shixiong Zhu},
  Charles Pritchard, Ankit Sardesai, Gideon Thomas, Daniel Berthereau, and
  Andreas Kohn.
\newblock {{OpenRefine}}.
\newblock 2019.

\bibitem{landau1989fast}
Gad~M Landau and Uzi Vishkin.
\newblock Fast parallel and serial approximate string matching.
\newblock {\em Journal of Algorithms}, 10(2):157--169, June 1989.

\bibitem{needleman1970general}
Saul~B. Needleman and Christian~D. Wunsch.
\newblock A general method applicable to the search for similarities in the
  amino acid sequence of two proteins.
\newblock {\em Journal of Molecular Biology}, 48(3):443--453, March 1970.

\bibitem{sarawagi2002interactive}
Sunita Sarawagi and Anuradha Bhamidipaty.
\newblock Interactive {{Deduplication}} using {{Active Learning}}.
\newblock page~10, 2002.

\bibitem{sariyar2010recordlinkage}
Murat Sariyar and Andreas Borg.
\newblock The {{RecordLinkage Package}}: {{Detecting Errors}} in {{Data}}.
\newblock 2:7, 2010.

\bibitem{tejada2002learning}
Sheila Tejada, Craig~A. Knoblock, and Steven Minton.
\newblock Learning domain-independent string transformation weights for high
  accuracy object identification.
\newblock In {\em Proceedings of the Eighth {{ACM SIGKDD}} International
  Conference on {{Knowledge}} Discovery and Data Mining - {{KDD}} '02}, page
  350, {Edmonton, Alberta, Canada}, 2002. {ACM Press}.

\end{thebibliography}

\begin{landscape}
  \begin{figure}
    \vspace{-2.5cm}
  \begin{tabular}{| p{3cm} | p{2.5cm} | p{2cm} | p{2.5cm} | p{2cm} | p{2cm} | p{2cm} | p{3cm} |}
    \hline
    \textbf{Name} & \textbf{Types} & \textbf{Properties} & \textbf{Retrieval} & \textbf{Name score} & \textbf{Property score} & \textbf{Popularity} & \textbf{Global score}  \\
    \hline
    \href{https://api.opencorporates.com/documentation/Open-Refine-Reconciliation-API}{OpenCorporates} & single (company) & {- jurisdiction\newline - date} & ElasticSearch and SQL \texttt{LIKE} on name & Combination of domain-specific rules and Levenshtein distance & Boolean on jurisdiction, fixed penalties on date & Fixed penalty on company types & Weighted sum of field scores \\
    \hline
    \href{http://data1.kew.org/reconciliation/help}{IPNI} & single (scientific name) & 19 properties & Lucene index & \multicolumn{2}{|p{6cm}|}{\href{https://github.com/RBGKew/Reconciliation-and-Matching-Framework/blob/master/rmf-configurations/src/main/resources/META-INF/spring/rmf-configurations/wcs-ipni-match/config_WCS_match0.xml}{Boolean matchers configured for each name and property, with canonicalization}} & N/A & \href{https://github.com/RBGKew/Reconciliation-and-Matching-Framework/blob/185968ad27bbe34c606c7c4cae0556399f295825/rmf-core/src/main/java/org/kew/rmf/core/lucene/LuceneUtils.java}{Average of field scores adjusted for blank fields}  \\
    \hline
    \href{https://findthatcharity.uk/}{FindThatCharity.uk} & single (charity) & None & ElasticSearch &  TF-IDF & N/A & Logarithmic income & ElasticSearch's score \\
    \hline
    \href{https://numishare.blogspot.com/2017/10/nomisma-launches-openrefine.html}{Nomisma} & 23 types & \href{https://github.com/nomisma/framework/blob/ab5ead062c39ed87b2b29e3c227d7dc3a1881198/solr-home/1.5/nomisma/conf/schema.xml}{8 properties} & Solr & TF-IDF & \href{https://github.com/nomisma/framework/blob/c56093431e0d76aabb316ae53cfd6fadf6f44596/ui/xslt/serializations/json/reconcile-query.xsl}{Boolean filters} & N/A & Solr's score \\
    \hline
    \href{http://refine.codefork.com/}{VIAF} & 5 types & None & VIAF search API & \href{https://github.com/codeforkjeff/conciliator/blob/master/src/main/java/com/codefork/refine/viaf/sources/VIAFSource.java}{Levenshtein distance ratio} & \multicolumn{2}{|c|}{N/A} & (Name score only) \\
    \hline
    \href{http://refine.codefork.com/}{OpenLibrary} & single (book) & Unspecified & OpenLibrary search API, concatenating property values to name & \multicolumn{3}{|c|}{N/A} & Constant (1.0)\\
    \hline
    \href{http://refine.codefork.com/}{ORCID} & single (person) & None & ORCID search API & TF-IDF (assumed) & \multicolumn{2}{|c|}{N/A} & \href{https://github.com/codeforkjeff/conciliator/blob/master/src/main/java/com/codefork/refine/orcid/OrcidParser.java}{``relevancy'' score returned by the search API} \\
    \hline
    \href{https://tools.wmflabs.org/openrefine-wikidata/}{Wikidata} & Wikidata items used with \href{https://www.wikidata.org/wiki/Property:P31}{instance of (P31)} and \href{https://www.wikidata.org/wiki/Property:P279}{subclass of (P279)} &
    All Wikidata properties (a few thousands) & Wikidata search APIs & Levenshtein-based fuzzy metric & Defined by property datatype & Order by Qid (tie-breaker) & Weighted average of field scores \\
    \hline
    \href{http://blog.lobid.org/2018/08/27/openrefine.html}{lobid-gnd} & 8 types & \href{https://d-nb.info/standards/elementset/gnd}{All properties from the GND ontology} & ElasticSearch, concatenating property values to name & \multicolumn{2}{|c|}{TF-IDF} & N/A & ElasticSearch's score \\
    \hline
    \href{https://godot.date/tools/openrefine}{GODOT} & single (person) & None  & Direct comparison to all records & Levenshtein-based fuzzy metric & \multicolumn{2}{|c|}{N/A} &  (Name score only)\\
    \hline
    \href{https://github.com/alephdata/aleph/blob/master/aleph/views/reconcile_api.py}{OCCRP} & 8 types & defined by the schema & ElasticSearch & TF-IDF & \multicolumn{2}{|c|}{N/A} & ElaticSearch's score \\
    \hline
  \end{tabular}
  \caption{Main characteristics of the surveyed reconciliation services}
  \label{fig:services-table}
\end{figure}
\end{landscape}

\appendix

\end{document}